\documentclass[fleqn,10pt]{wlscirep}
\usepackage[utf8]{inputenc}
\usepackage[T1]{fontenc}
\usepackage{tabularx}

\title{Impossible by Conventional Means: Ten Years on from the DARPA Red Balloon Challenge}

\author[1,+,*]{Alex Rutherford}
\author[1,+,*]{Manuel Cebrian}
\author[1]{Inho Hong}
\author[1]{Iyad Rahwan}
\affil[1]{Center for Humans and Machines, Max Planck Institute for Human Development, Berlin, Germany}

\affil[*]{To whom correspondence should be addressed: rutherford@mpib-berlin.mpg.de, cebrian@mpib-berlin.mpg.de}

\affil[+]{These authors contributed equally to this work}

\begin{abstract}
Ten years ago, DARPA launched the ‘Network Challenge’, more commonly known as the ‘DARPA Red Balloon Challenge’. Ten red weather balloons were fixed at unknown locations in the US. An open challenge was launched to locate all ten, the first to do so would be declared the winner receiving a cash prize. A team from MIT Media Lab was able to locate them all within 9 hours using social media and a novel reward scheme that rewarded viral recruitment~\cite{cebrian2012finding}. This achievement was rightly seen as proof of the remarkable ability of social media, then relatively nascent, to solve real world problems such as large-scale spatial search. Upon reflection, however, the challenge was also remarkable as it succeeded despite many efforts to provide false information on the location of the balloons. At the time the false reports were filtered based on manual inspection of visual proof and comparing the IP addresses of those reporting with the purported coordinates of the balloons. In the ten years since, misinformation on social media has grown in prevalence and sophistication to be one of the defining social issues of our time. Seen differently we can cast the misinformation observed in the Red Balloon Challenge, and unexpected adverse effects in other social mobilisation challenges subsequently, not as bugs but as essential features. We further investigate the role of the increasing levels of political polarisation in modulating social mobilisation. We confirm that polarisation not only impedes the overall success of mobilisation, but also leads to a low reachability to oppositely polarised states, significantly hampering recruitment. We find that diversifying geographic pathways of social influence are key to circumvent barriers of political mobilisation and can boost the success of new open challenges.
\end{abstract}
\begin{document}

\flushbottom
\maketitle

\thispagestyle{empty}

\section*{The DARPA Red Balloon Challenge and its Progeny}

The DARPA Network Challenge began a series of open challenges (see Table 1) that explored different facets of social mobilisation. The challenge conditions were able to incentivise researchers to focus on particular problems in sectors as diverse as intelligence, health and problem solving with the lure of prestige or financial reward. In each case the time constraints and high-profile nature of the challenge pushed the parameters of the task to extremes uncovering sometimes unsettling adversarial effects. 

\begin{table}[h]
\centering
\fontfamily{phv}\selectfont
\begin{tabular}{ |c|c|c|p{6.7cm}|} 
    \hline
    \textbf{Challenge} & \textbf{Year} & \textbf{Theme} & \textbf{Unexpected Lesson} \\ 
    \hline
    DARPA Red Network Challenge & 2009 & Social Search & Misinformation is inevitable~\cite{cebrian2012finding,tang2011reflecting,pickard2011time,chen2016bandit}\\ 
    \hline
    DARPA Shredder Challenge & 2011 & Problem Solving & Sabotage has asymmetric power\cite{stefanovitch2014error,naroditskiy2014crowdsourcing,oishi2014iterated}\\
    \hline
    Nexus 7 (More Eyes) & 2011 & Intelligence & Privacy concerns restrict uptake\cite{weinberger2017imagineers}\\
    \hline
    My HeartMap & 2011 & Health & Social media needs mass Media~\cite{heartmap}\\
    \hline
    State Department Tag Challenge & 2012 & Social Search & Misinformation and sabotage are inevitable~\cite{rutherford2013targeted, rahwan2012global,naroditskiy2012verification}\\
    \hline
    CLIQR Quest challenge & 2012 & Social Search & Social media needs mass media \& \newline Misinformation and sabotage are inevitable~\cite{cliqr}\\
    \hline
    Langley Castle Challenge & 2014 & Social Search & More similar (homophilous) friends mobilize faster~\cite{alstott2014homophily,wang2015effect}\\
    \hline
    WeHealth & 2015 & Health & Privacy concerns restrict uptake~\cite{wehealth}\\
    \hline
    FiftyNifty & 2017 & Political Mobilisation & Political polarisation undermines bi-partisan mobilization~\cite{fiftynifty}\\
    \hline
    Black Rock Atlas & 2018 & Social Search & Social media needs mass media~\cite{epstein2019towards}\\
    \hline
\end{tabular}
\caption{Open challenges and the lessons.}
\end{table}

Following the Red Balloon Challenge, DARPA sponsored the Shredder Challenge to understand vulnerabilities in the crowd-sourced reconstruction of fragments of shredded documents. A team led by the member of the MIT team which won the DARPA Network Challenge (by then at the University of California, San Diego) was the top crowdsourcing solution, yet failed to complete the challenge, partly due to a coordinated series of attacks that sabotaged the progress made by other contributors~\cite{stefanovitch2014error}.

One year later the US State Department sponsored the Tag Challenge, in which teams were invited to locate 5 mobile human `targets’ in 5 different cities on a given day~\cite{rutherford2013targeted, rahwan2012global}. This challenge was won by a team led from Masdar Institute in the UAE in collaboration with same member of the MIT team which won the DARPA Network Challenge. Despite its success, the team’s progress was hindered by several adversarial activities such as sabotage of the team’s platform and impersonation of the team’s identity on social media.

\begin{figure}[t!]
    \centering
    \includegraphics[width=\linewidth]{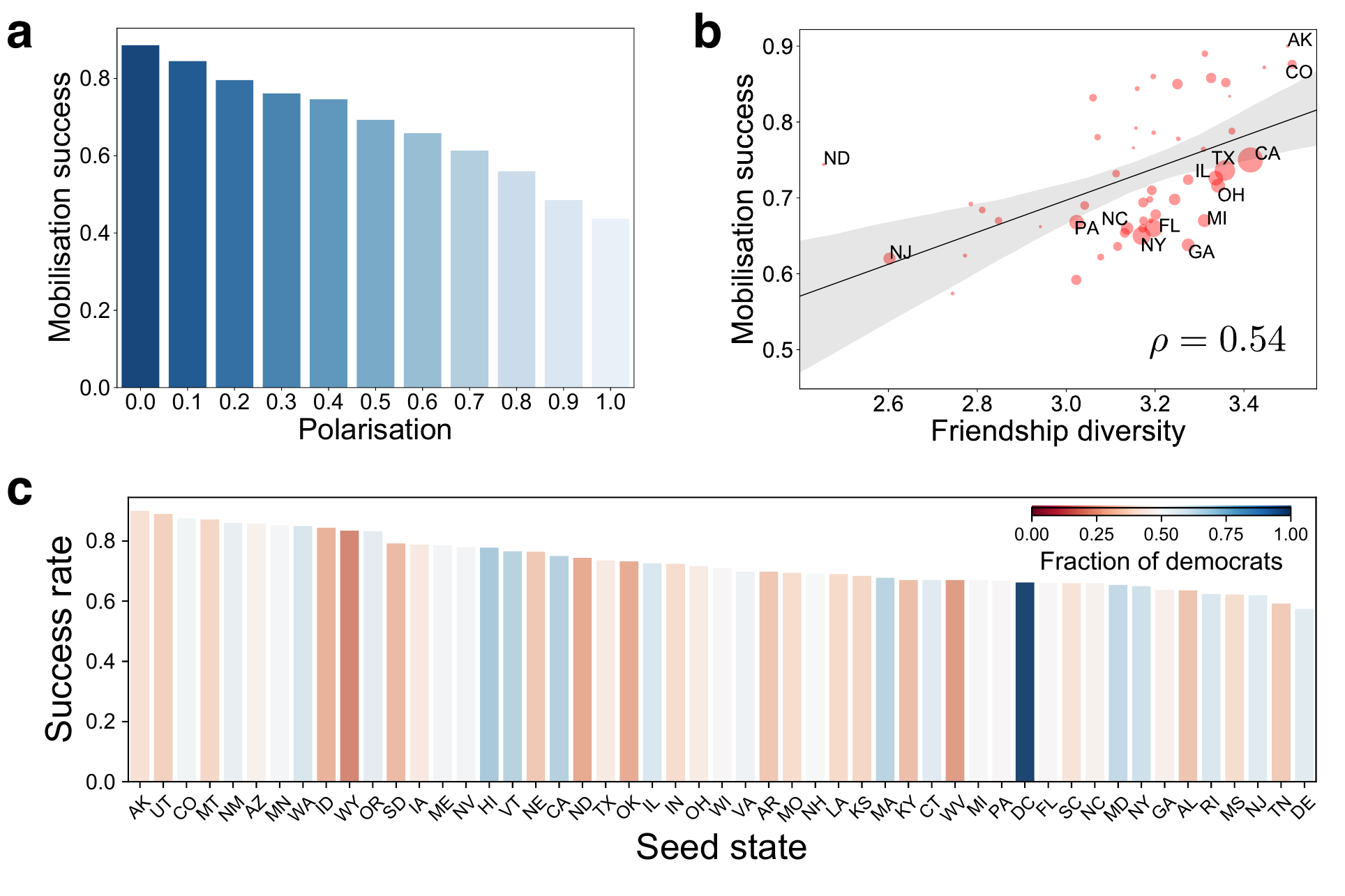}
    \caption{(a) Rate of success in the simulation as a function of political polarisation. Our simulation reconstructs the mobilisation process in the FiftyNifty challenge. The success rate is calculated from 1,000 simulations for each polarisation parameter $\alpha$. (b) Friendship diversity and rate of success of seed states. The diversity measured from the entropy of friendship weights to the other states shows high correlation (i.e., $\rho=0.54$) with the success rate of mobilisation from the seed state. (c) Size of mobilisation for different seed states for moderate polarisation (i.e., $\alpha=0.5$). The colours denote the political makeup of each seed state (blue for Democratic and red for Republican). See the Supplementary Information for details on the simulations.}
    \label{fig:polarisation}
\end{figure}

In 2017, the Viral Communications research group in MIT Media Lab launched the FiftyNifty challenge~\cite{fiftynifty}. The objective of which was to mobilise voters in each of the 50 states to contact their representatives through viral recruitment. Building on the successful modeling of social mobilisation phenomena~\cite{rutherford2013limits} we show that the success of such a recruitment process is susceptible to political polarisation. A higher degree of polarisation not only decreased the overall success rate of the challenge (Fig.~\ref{fig:polarisation}a) but also made it harder to reach oppositely polarised states (see Figs.~\ref{fig:call} and~\ref{fig:net_call} in the Supplementary Information).

These adversarial activities were unexpected outcomes of these challenges, but they were surprisingly prescient of modern day adversarial activities using social media and digital platforms. While it might be tempting to favour fewer such challenges that might encourage such side-effects, we conclude the opposite. These examples have shown that we need more of these challenges to bring possible side effects to prominence\cite{naroditskiy2014crowdsourcing,oishi2014iterated}. Given the unprecedented complexity of AI and human ecologies, we see a need for open challenges focusing on deployment of AI systems into society. This is consistent with recent calls for a dedicated study of the sociology of AI agents~\cite{rahwan2019machine}. 

\subsection*{Mobilisation impeded by political polarisation}

Based on the negative experience in the Fifty-Nifty challenge, we now investigate the effect of polarisation on mobilisation. Political polarisation was not present at the current level for early challenges, so it is important to understand how it could affect future ones. For that, we simulate a social mobilisation process to show how political polarisation of individuals may impede mobilisation over the entire country. In this simulation, we show (1) the success rate impeded by polarisation, (2) the reachability to each state by polarisation and (3) good and bad seed states for mobilisation.

We reconstruct mobilisation in the FiftyNifty challenge by simulating the branching recruitment process~\cite{rutherford2013limits} that spreads through the county-wise friendship network~\cite{bailey2018social} given by the Facebook Social Connectedness Index (SCI) dataset on the political landscape of the Unities States~\cite{election2016}.
This novel dataset of Facebook SCI represents the normalized counts of friendship pairs between the entire US counties, and the political landscape is based on the result of the 2016 Presidential Election.
In the simulation, a political polarisation parameter $\alpha$ ranging from 0 (no polarization) to 1 (full polarization) controls the success probability of recruits between a recruiter and a recruited as $p=1-\alpha$ for opposite polarisation in contrast to $p=1$ for identical polarisation. Thus, polarisation only affects recruitment while the friendship network is fixed.

We start by first measuring the success rate of mobilisation for different levels of polarisation. In the simulation, mobilisation starts from 1,000 Democratic seeds in Middlesex County, MA (the location of Cambridge, MA from where the challenge was launched). We measure the success rate for each polarisation parameter $\alpha$ where some level of mobilisation in each state is required for success. As a result, we observe a monotonic decrease of the mobilisation success with increasing polarisation (see Fig.~\ref{fig:polarisation}a). This finding shows that political polarisation between individuals hampers mobilisation across different regions.

\subsubsection*{The hardest place to reach}

So far, we observe the overall decrease of the success rate by polarisation. Then, how does polarisation affect mobilisation over different states? We first check the relation of the mobilisation size and the population size of states. Fig.~\ref{sfig:polarisation} shows a high correlation between mobilisation and population. Political polarisation decreases the mobilisation size in every state. Then, does polarisation bring equal impacts to states?

To see the polarisation effect in different regions, we compare the mobilisation size of each state with its political makeup. Fig.~\ref{fig:call} shows that Democratic states have a larger mobilisation size compared to Republican states in general. Therefore, both polarisation and population affect the success of mobilisation. This combination leads to a low reachability to small and Republican state; For example, Wyoming is the hardest place to reach for a campaign started from Democrats in Massachusetts.

By decomposing the population trend from the mobilisation size, we show the effect of polarisation on the success of mobilisation in each state. We subtract the population trend in Fig.~\ref{sfig:polarisation} from the mobilisation size. As a result, Fig.~\ref{fig:net_call} shows a clear separation between states in the mobilisation size by the political makeup. Also, higher polarisation makes a clearer separation between blue and red states. Therefore, it confirms that polarisation not only impedes the overall success of mobilisation, but also leads to a low reachability from the opposite political spectrum.

\subsubsection*{What determines a good seed state?}

The simulation so far was focused on reproducing the mobilisation process in the FiftyNifty challenge seeded from Middlesex County, MA. If we launch a new challenge or campaign using the polarised friendship network, where is the best seed state for its success? To answer this question, we simulate the mobilisation process seeded from each state in turn. We assume that Democratic seeds are located in the most populated county of each state. Fig~\ref{fig:polarisation}c shows the success rate of mobilisation from different seed states; Alaska is the best seed state while Delaware is the worst seed state. The variability of seed quality is not determined by the political makeup of states.

Then, why are some states better than the others in seeding the mobilisation process? We find the answer in the connection diversity of each seed state. We define the friendship diversity of each state using the entropy of connection probability to other states as
\begin{equation}
    H_{i} = -\sum_{j}p_{ij}\log{p_{ij}}
    \label{eq:diversity}
\end{equation}
where $p_{ij}=w_{ij}/\sum_{j}w_{ij}$ is the connection probability from state $i$ to state $j$, and $w_{ij}$ is the aggregated weight of friendship on Facebook.
As a result, Fig.~\ref{fig:polarisation}b shows a high correlation (i.e., $\rho=0.54$ with $p < 10^{-4}$) between connection diversity $H_{i}$ of seed state $i$ and its success rate of mobilisation. 
This finding gives us two implications for successful mobilisation. First, the location of seeds is important in mobilisation. As the mean branching factor is less than 1 (i.e., the number of new recruits by a recruiter is less than 1 on average.), the depth of the recruitment network is finite. Thus, the first generation of recruits in the recruitment network plays a key role in overall mobilisation. Accordingly, the connection diversity of a seed state controls the spread of mobilisation to other states through a friendship network. When a seed state is evenly connected to the other states, mobilisation is easy to spread to many states. On the contrary, if a seed state has strongly biased connections to a few states, these states would take up most of the outgoing recruitment flows from the seed state. This finding demonstrates that diversification of the pathways of social influence would be the key to the success of new open challenges.\\

\section*{Societal Testing}

The allure of open challenges is clear when considering how other new technologies are developed, tested and launched. In all cases, tests are first performed in a controlled environment and subsequently the test environment is gradually enlarged to successively model a real world deployment more closely. For example, software is unit tested, integration tested and system level tested, and subsequently, when in production, is subjected to user-driven bug reports and hacker-driven penetration testing. Likewise, new drugs may be tested on animals, small scale medical trials followed by open usage among certain demographics. 


For technologies such as drugs, regulation is well developed and the procedure for release is well understood. Thus following medical trials, once it is established with reasonable certainty and noting reasonable caveats, that the drug’s benefits outweigh its harms when properly administered, it is freely released for usage within society with strict usage instructions. The majority of testing is \textit{pre-deployment} and limited testing is done \textit{post-deployment}, for example looking at longitudinal effects or interactions with other drugs.

\begin{figure}[ht]
    \centering
    \vspace{0.3cm}
    \includegraphics[width=\linewidth]{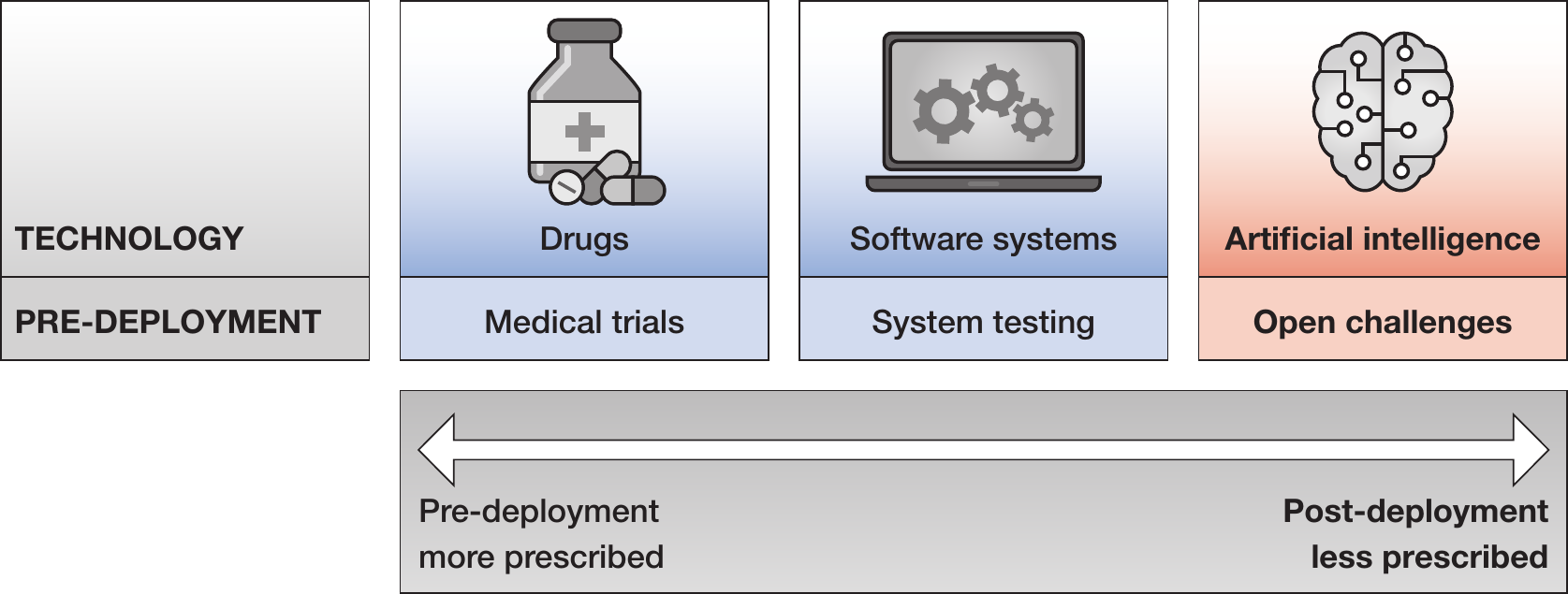}
    \caption{A prescriptive spectrum of technologies and testing methodologies.}
    \vspace{0.2cm}
    \label{fig:spectrum}
\end{figure}

For software systems, there exists slightly more balance between \textit{pre-} and \textit{post-deployment} testing. Once technical functionality has been tested and established, user testing takes place to see if the intended functionality persists under realistic usage patterns in a trial deployment. Following deployment into production, testing continues through continuous telemetry, A/B tests of new features and vulnerability tests through bug bounties and sanctioned white hat hacking. The reason why post-production testing is more thorough, is that software is an example of a socio-technical system~\cite{vespignani2012modelling}. Typically of such systems, the behaviour of computer software is determined by the coupled behaviour of its users and groups of users, in a non-trivial manner and such that testing of its behaviour cannot be conducted in isolation and without the realistic usage patterns of the user.

Such is the complexity and versatility of social media platforms and other Internet mediated technologies, the expected behaviour of e.g. many-to-many instant messaging, Virtual Reality or DeepFakes cannot simply be interpolated from a small subsample, or a limited user testing. While some might object to open challenges as being reckless and unsanctioned societal experimentation, we can ask what are the alternatives? Internet platform companies have resorted to closed focus groups followed by the global release of largely unregulated tools in an arms race of engagement; a process that has been described as ‘society as a beta test’.

Reflecting on 10 years of social media challenges, we argue that emerging technologies represent socio-technical systems of such complexity and with such far reaching effects that suitable testing should emphasise post-deployment testing as much as \textit{pre-deployment} testing. The challenges discussed above can be considered as a forum for transparent, white hat attacks on these platforms operating at the limits of their parameters. Below we reflect upon the common characteristics of successful challenges.

\begin{itemize}
\item \textbf{Loosely defined and possibly tangential}: The Red Balloon Challenge was not explicitly about social media, yet this emerged as the most appropriate medium. It is also important that challenges are sufficiently loosely defined to allow the system freedom to unearth novel behaviours.
\item \textbf{Non-prescriptive}: Challenge platforms such as Kaggle incentivise incremental improvements in cases when we have converged upon a working solution. Many of the challenges described above failed or only partially succeeded.
\item \textbf{Failure is likely (and fine)}: following from the previous point, the challenge should shift emphasis from successfully (or not) completing the stated objective, but uncovering and reporting unexpected side-effects and adversarial efforts.
\item \textbf{Has low barrier to entry}: to maximise the freedom for solutions to explore the far reaches of the parameter space challenges should not impose onerous conditions or an entry fee. Diverse connectivity through open participation is the key to successful mobilisation (see Fig.~\ref{fig:polarisation}b).
\item \textbf{Public}: challenges should allow for public and transparent publications of findings to enable society to engage with the potential risks and benefits. For example, the Red Balloon challenge findings were published in \textit{Science} and \textit{PNAS} as well as receiving publicity on \textit{The Colbert Show} and in \textit{The New York Times}.
\item \textbf{Prestigious}: Successful challenges should be sanctioned by prestigious public institutions such as DARPA or the State Department. This encourages uptake and participation of competitors, users and adversaries.
\end{itemize}

These challenges that followed the DARPA balloon challenge brought to light some unsavoury faces of Internet mediated technologies. Aspects that have become all too familiar: sabotage, polarisation and misinformation. Yet we are in a better place having illuminated and characterised these ‘edge cases’, as they were considered at the time, and better placed to deal with them now that they are mainstream. Just as we began to understand and characterise the power and pitfalls of social media with the Red Balloon Challenge a decade ago, we call urgently for a further decade of challenges investigating AI as it emerges as a mainstream and general purpose technology.

\section*{Contributions}

M.C. conceived the study. M.C., A.R. designed the study. A.R., I.H. developed the computational model. I.H. performed simulations. M.C., A.R., I.H. analyzed data. A.R. wrote the manuscript. M.C., A.R., I.R., I.H. edited the manuscript. 

\section*{Acknowledgements}

We are grateful to May Al-Hazzani for early investigations of the FiftyNifty data, to Andrew Lippmann, Léopold Mebazaa, Travis Rich, Jasmin Rubinovitz, Simon Johnson, Jason Seibel and Penny Webb for the dataset, and to J\"urgen Rossbach for graphics editing.

\bibliography{reference}

\newpage
\section*{Supplementary Information}

\setcounter{figure}{0}
\setcounter{equation}{0}
\setcounter{section}{0}
\renewcommand{\theequation}{S\arabic{equation}}
\renewcommand{\thefigure}{S\arabic{figure}}

\subsection*{Simulating polarised mobilisation}

The overall mobilisation process adopts the branching recruitment process in Rutherford \textit{et al}~\cite{rutherford2013limits}. Starting from a group of seed recruiters, mobilisation occurs repeatedly over a social network until a goal of mobilisation is reached (i.e., success in open challenges) or there is no more recruits. The prominent improvements of our simulation model from the original mobilisation model are (1) the implement of political polarisation modeled from literature~\cite{bakshy2015exposure}, (2) an empirical social network from the Facebook Social Connectedness Index (SCI) dataset~\cite{bailey2018social} and (3) an empirical political landscape from the 2016 US Presidential Election~\cite{election2016}. The following description shows how these improvements are implemented in our simulation process of seeding, activation, recruitment and termination.\\

\noindent\textbf{Seeding.} 
The simulation starts from $N_s$ Democratic or Republican individual seeds in a specific county of US. The simulation for the FiftyNifty challenge is seeded from Middlesex County (the location of Cambridge, MA from where the challenge was launched). When we simulate mobilisation for different seed states, we chose the most populated county of each state as the seed counties.
The activation time $\Delta t_a(i)$ of seed $i$ is determined by a log-normal distribution $P(\Delta t_a)$ with a mean of 1.5 day and a standard deviation of 5.5 days~\cite{rutherford2013limits}. The individual information is inserted into a priority queue, and is drawn one-by-one in the order of the shortest activation time.
Also, the number of friends $k$ (i.e., branching factor) for recruitment is assigned to each individual seed following a Harris discrete distribution~\cite{rutherford2013limits} $P(k)$ as,
\begin{equation}
    P(k) = \frac{H_{ab}}{b + k^{a}},
    \label{eq:branching}
\end{equation}
where $a$ is a power-law exponent, $b$ is fitted to a given empirical mean value of branching factors, and $H_{ab}$ is a normalisation factor. In the simulation, we use $a=2.1$ and $\langle k \rangle=0.9$~\cite{rutherford2013limits}.\\

\noindent\textbf{Activation.}
In each simulation step, one individual is drawn from the queue in the order of the shortest activation time. This activated individual is mobilised with a probability of 1 if it has identical polarisation with its recruiter. In the case of opposite polarisation with its recruiter, the mobilisation probability is given as $1-\alpha$ where $\alpha$ is the polarisation parameter ranging from 0 to 1. The mobilisation probability of seeds is given to 1. If mobilisation is successful, the activated individual further recruits its friends by the following recruitment process.\\

\noindent\textbf{Recruitment.}
A mobilised individual recruits $k$ friends following the branching factor $k$ which was assigned in the previous recruitment or seeding. Each friend has 4 stochastically chosen properties: activation time, branching factor, residence county and the political orientation. 
\begin{itemize}
\item The waiting time for activation $\Delta t_a$ is chosen by the log-normal distribution in \textbf{Seeding}. If the current time is $t$, the recruited friend is actiavated at $t + \Delta t_a$.

\item The branching factor is determined by the Harris distribution in Eq.~\eqref{eq:branching}. As the mean of branching factors is less than one, the branching recruitment process terminates eventually.

\item The residence county of each friend is determined by the strength of friendship given by the Facebook dataset. If a recruiter is in county $i$, the probability of recruiting a friend in county $j$ is $p_{ij} = w_{ij}/\sum_{j}w_{ij}$, where $w_{ij}$ is the Social Connectedness Index (SCI) that represents the number of friend pairs in the Facebook dataset~\cite{bailey2018social}. As the goal of the FiftyNifty challenge was to mobilize every state, recruits within the same state are prohibited in the simulation.

\item The political orientation of a recruited friend is chosen by the combination of the political makeup of its residence county and polarisation. Political polarisation between friends is known to be identical with a probability of $3/4$ and opposite with $1/4$ by homophilic friendship~\cite{bakshy2015exposure}. Also, the probability of the political orientation is proportional to the political makeup of the residence county given by the 2016 US Presidential Election~\cite{election2016}. Combining them leads to
\begin{equation}
(p_{dem}, p_{rep}) \sim \begin{cases}
(\frac{3}{4}pol_{j}, \frac{1}{4}(1-pol_{j})) &\text{if the recruiter is Democratic},\\
(\frac{1}{4}pol_{j}, \frac{3}{4}(1-pol_{j})) &\text{if the recruiter is Republican},
\end{cases}
\end{equation}
where $p_{dem} = 1-p_{rep}$ is the probability of being Democratic and $pol_{j}$ is the proportion of votes to the Democratic party in the 2016 US Presidential Election in residence county $j$ of the recruited.\\
\end{itemize}

\noindent\textbf{Termination.}
The simulation terminates when we observe at least one mobilised individual in every state (i.e., ``Success''), or there is no remaining individual to be activated in the queue (i.e., ``Failure'').\\

\begin{figure}[bth]
    \centering
    \includegraphics[width=0.6\linewidth]{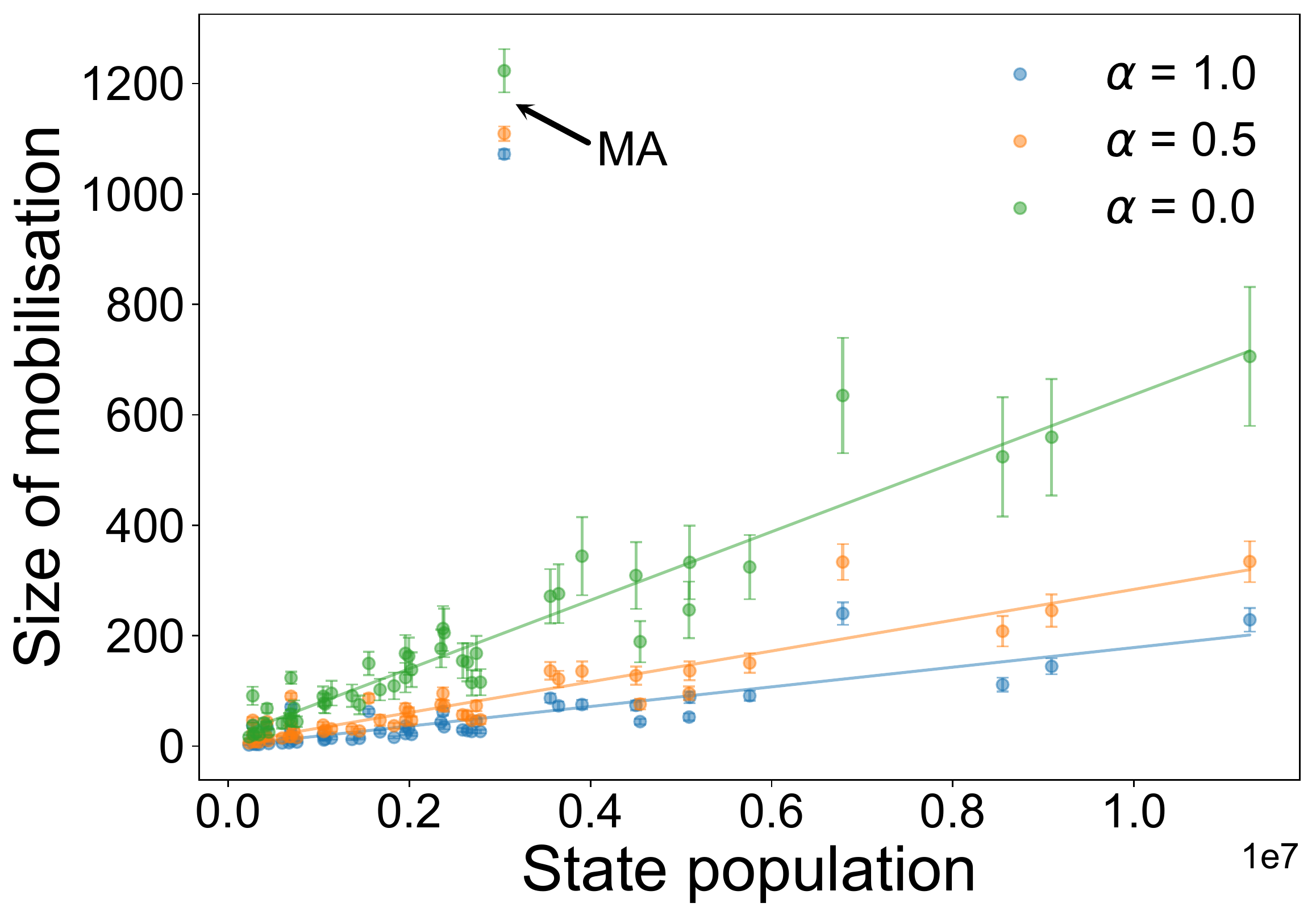}
    \caption{Size of mobilisation as a function of population size for different levels of polarisation ($\alpha=\{0.0, 0.5, 1.0\}$). The result is the average of 500 simulations for 1,000 Democratic seeds in Middlesex County, MA. The error bars denote the 95\% interval of the mobilisation size.}
    \label{sfig:polarisation}
\end{figure}

\begin{figure}[phtb]
    \centering
    \includegraphics[width=0.8\linewidth]{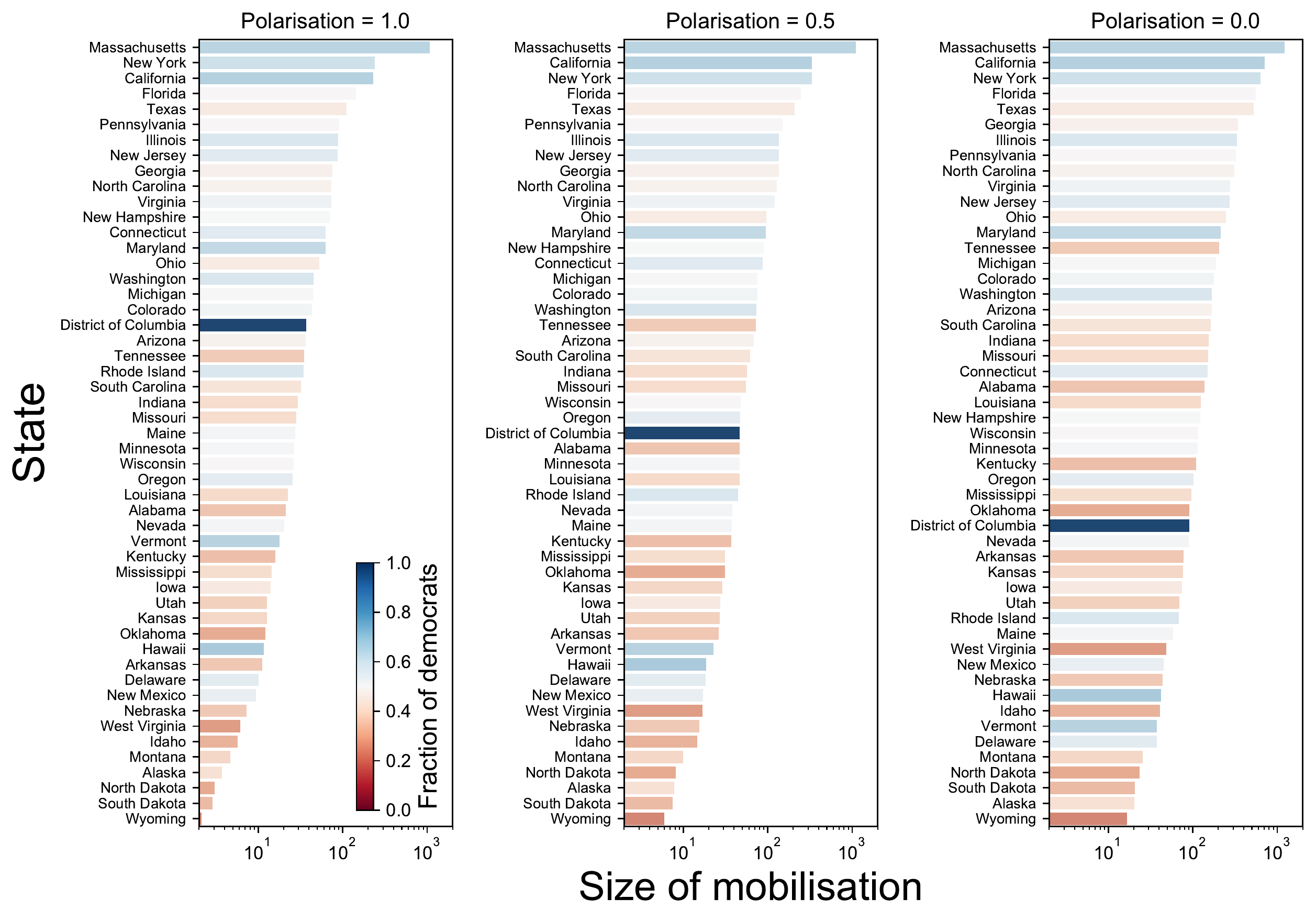}
    \caption{Size of mobilisation in each state for different levels of polarisation ($\alpha=1.0, 0.5, 0.0$ (left to right). The result is the average of 500 simulations for 1,000 Democratic seeds in Middlesex County, MA. The colour denote the political makeup of each state (blue for Democratic and red for Republican).}
    \label{fig:call}
\end{figure}

\begin{figure}[phtb]
    \centering
    \includegraphics[width=0.8\linewidth]{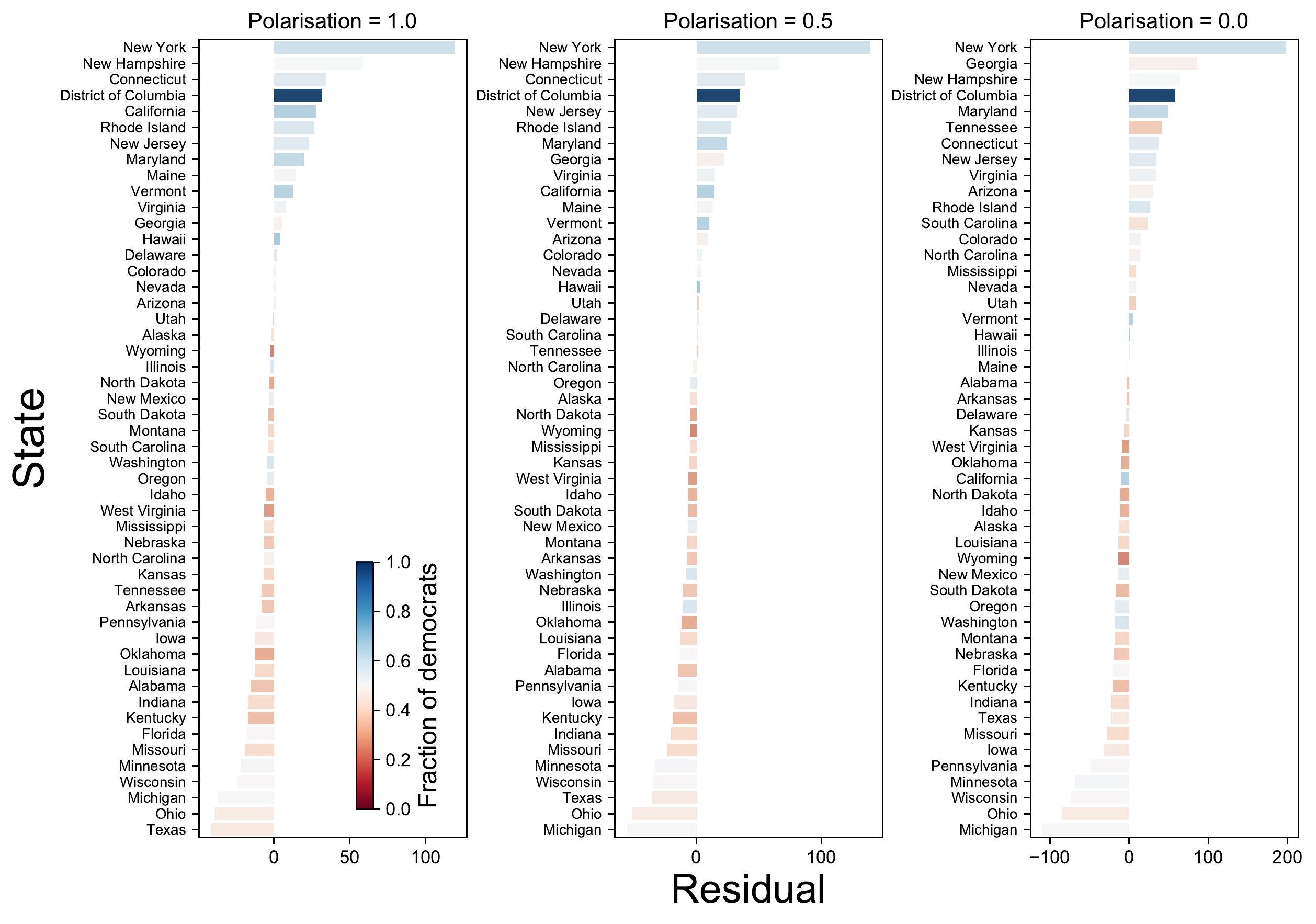}
    \caption{Size of mobilisation without population trends for different levels of polarisation ($\alpha=1.0, 0.5, 0.0$ (left to right). The result is the average of 500 simulations for 1,000 Democratic seeds in Middlesex County, MA. (blue for Democratic and red for Republican).}
    \label{fig:net_call}
\end{figure}

\end{document}